\documentclass{aa}
\usepackage[varg]{txfonts}

\usepackage[colorlinks=true,citecolor=blue,linkcolor=blue,bookmarks=false]{hyperref}

\usepackage{booktabs}
\usepackage{natbib}
\bibpunct{(}{)}{;}{a}{}{,}
\usepackage{supertabular}
\usepackage{array}
\usepackage{float}
\usepackage{varioref}
\usepackage{multirow}
\usepackage{graphicx}
\usepackage{epstopdf}
\usepackage{longtable}
\usepackage{comment}
\usepackage{threeparttable, tablefootnote}
\usepackage[T1]{fontenc}
\usepackage{etoolbox}
\usepackage{siunitx}
\usepackage{caption}
\usepackage{subcaption}
\usepackage{tabu}
\usepackage{tikz-cd}
\DeclareGraphicsExtensions{.eps}
\defcitealias{chapman2005}{Chapman2005}

\begin{document}

\title{\LARGE Searching for obscured AGN in $z \sim$ 2 submillimetre galaxies}


\author{H. Chen\inst{1, 2, 3}
\and M. A. Garrett\inst{2, 4} 
\and S. Chi\inst{5, 6, 7}
\and A.~P. Thomson\inst{2} 
\and P. D. Barthel\inst{5} 
\and D. M. Alexander\inst{8} 
\and T.~W.~B. Muxlow\inst{2} 
\and R.~J. Beswick\inst{2} 
\and J.~F. Radcliffe\inst{2, 9, 10} 
\and N.~H. Wrigley\inst{2} 
\and D. Guidetti\inst{11} 
\and M. Bondi\inst{11}
\and I. Prandoni\inst{11} 
\and I. Smail\inst{8} 
\and I. McHardy\inst{12} 
\and M.~K. Argo\inst{2, 13} 
} 
     

\institute{Shanghai Astronomical Observatory, 80 Nandan Road, Xuhui District, Shanghai, 200030, China
\and Jodrell Bank Centre for Astrophysics (JBCA), Department of Physics \& Astronomy, Alan Turing Building, The University of Manchester, M13 9PL, United Kingdom 
\and University of Chinese Academy of Sciences, 19A Yuquanlu, Beijing 100049, China
\and Leiden Observatory, Leiden University, PO Box 9513, 2300 RA Leiden, The Netherlands
\and Kapteyn Astronomical Institute, University of Groningen, PO Box 800, 9700 AV Groningen, The Netherlands
\and Joint Institute for VLBI in Europe (JIVE), PO Box 2, 7990 AA Dwingeloo, The Netherlands
\and Netherlands Foundation for Research in Astronomy (ASTRON), PO Box 2, 7990 AA Dwingeloo, The Netherlands
\and Centre for Extragalactic Astronomy, Department of Physics, Durham University, South Road, Durham DH1 3LE, UK
\and Department of Physics, University of Pretoria, Lynnwood Road, Hatfield, Pretoria, 0083, South Africa
\and South African Radio Astronomy Observatory, 3rd Floor, The Park, Park Road, Pinelands, Cape Town, 7405, South Africa
\and INAF \text{-} Istituto di Radioastronomia, via Gobetti 101, I-40129 Bologna, Italy
\and Physics and Astronomy, University of Southampton, Southampton SO17 1BJ, UK
\and Jeremiah Horrocks Institute, University of Central Lancashire, Preston PR1 2HE, UK
} 
 
\date{Revised April 16, 2020 / Accepted April 10, 2020}

\abstract
{}
{Submillimetre-selected galaxies (SMGs) at high redshift ($z$ $\sim$ 2) are potential host galaxies of active galactic nuclei (AGN). If the local Universe is a good guide, $\sim$ 50$\%$ of the obscured AGN amongst the SMG population could be missed even in the deepest X-ray surveys. Radio observations are insensitive to obscuration; therefore, very long baseline interferometry (VLBI) can be used as a tool to identify AGN in obscured systems. A well-established upper limit to the brightness temperature of 10$^5$ K exists in star-forming systems, thus VLBI observations can distinguish AGN from star-forming systems via brightness temperature measurements.}
{We present 1.6 GHz European VLBI Network (EVN) observations of four SMGs (with measured redshifts) to search for evidence of compact radio components associated with AGN cores. For two of the sources, e-MERLIN images are also presented.}
{Out of the four SMGs observed, we detect one source, J123555.14, that has an integrated EVN flux density of 201 $\pm$ 15.2 $\mu$Jy, corresponding to a brightness temperature of 5.2 $\pm$ 0.7 $\times$ 10$^5$ K. We therefore identify that the radio emission from J123555.14 is associated with an AGN. We do not detect compact radio emission from a possible AGN in the remaining sources (J123600.10, J131225.73, and J163650.43). In the case of J131225.73, this is particularly surprising, and the data suggest that this may be an extended, jet-dominated AGN that is resolved by VLBI. Since the morphology of the faint radio source population is still largely unknown at these scales, it is possible that with a $\sim$ 10 mas resolution, VLBI misses (or resolves) many radio AGN extended on kiloparsec scales.}
{}

\titlerunning{Searching for Obscured AGN in z $\sim$ 2 SMGs}
\authorrunning{H. Chen et al.(2019)}
\maketitle

\section{Introduction} \label{1}

Submillimetre galaxies (SMGs) are the most bolometrically luminous sources in the Universe \citep{swinbank2014}. They are responsible for up to half of the total star formation in the Universe and are likely the massive, dusty progenitors of the largest elliptical galaxies that we see in the local Universe (e.g. \citealt{simpson2014, swinbank2006, casey2014}). While it is clear that the processes of star formation are important in SMGs, it is still unclear as to what fraction also hosts active galactic nuclei (AGN) and whether those AGNs are energetically significant in the bolometric output of those luminous galaxies. Some SMG systems probably host components associated with both AGN activity and star-formation processes (for a review, see \citealt{biggs2010}). Previous observations of extragalactic objects have illustrated that the upper limit to the brightness temperature of $z~>~0.1$ star-forming galaxies is expected not to exceed $T_b$ $\sim 10^{5}$ K, and distant SMGs with $T_b$ above this value are most likely powered by AGN (e.g. \citealt{condon1992, middelberg2010, middelberg2013}).

To date, large surveys focused on detecting AGN systems have been conducted at a range of different wavelengths and, in particular, for the IR and X-ray domains \citep{alexander2005, alexander2008, ivison2004, lutz2005, MD2007, MD2009, valiante2007, pope2008, bonzini2013, smolcic2017, wang2013, stach2019}. Due to the dust extinction in the near-IR, optical, and UV, as well as gas absorption in X-ray bands, these surveys are often incomplete. In the mid- and far-IR, incompleteness of AGN surveys may arise from the fact that not all AGN have significant IR emission from a dusty torus and therefore may not be detected. It is postulated that up to a third of AGN are undetected in these surveys \citep{mateos2017}. Moreover, other studies have compared AGN selected from various wavebands and find that their host galaxies tend to have different properties in terms of colour \citep{hickox2009} and star-formation rates (SFR) \citep{juneau2013, ellison2016}. In particular, \citet{hickox2009} illustrated that there is only very little overlap between their 122 radio-selected AGN and those selected by X-ray or IR. Therefore, dust-free radio surveys are needed to provided a more complete census of the AGN population. Traditional radio surveys are only sensitive to radio-loud (RL) AGN, which only represent a tiny fraction (10 $\sim$ 20\%) of the whole AGN population; however, modern radio surveys can achieve a flux depth where radio-quiet AGN can be detected (see \citealt{Prandoni2018} for a review). Recent work has focused on the radio as it is sensitive to AGN and star formation concordantly, thus providing a method of surveying AGN and star-formation activity across cosmic time (e.g. \citealt{smolcic2017, padovani2015}). A lot of work has also been done to look for AGN-driven radio emission, which has been identified by an excess of radio emission compared to what is expected based on the radio-FIR correlation, holding for star-forming galaxies (e.g. \citealt{ivison2010, condon2002, thomson2014, magnelli2015}). 

Very long baseline interferometry (VLBI) provides an alternative method of identifying AGN that is not affected by star-formation-related radio emission via the detection of high brightness temperature compact radio components as can be seen in \citet{garrett2001, garrett2005, chi2013, middelberg2011, middelberg2013, HR2017}, for example. In particular, \citet{chi2013} identified 12 AGNs in the \textit{Hubble} Deep Field-North (HDF-N) using a global array of VLBI telescopes. As suggested in \citet{chi2009}, the radio-enhanced AGNs are probably obscured AGNs which remain undetected even in the deepest X-ray surveys. Indeed, if the local Universe is a good guide, roughly half of the AGNs in SMGs are Compton thick (e.g.  \citet{risaliti1999}). 
 
\citet{chi2009} argued that deep, high-resolution radio observations are required in order to generate complete samples of obscured AGNs at high redshifts. However until recently, deep, wide-field VLBI surveys were difficult to realise and the field of view (FoV) is still relatively limited. The employment of new analysis techniques, as seen in \citet{radcliffe2016} and \citet{deller2011} for example, has permitted much deeper and wider VLBI surveys of AGN to be conducted (e.g. \citealt{middelberg2013, HR2017, radcliffe2018}). VLBI has become a sensitive tool in distinguishing between AGN and star-formation processes, including the ability to detect Compton-thick AGNs in dusty systems that would otherwise go undetected. These advances have proven that the radio morphological information and brightness temperature measurements provided by VLBI are sensitive enough to isolate the pure star-forming regions from AGN within individual SMGs (e.g. \citealt{muxlow2005, chi2013}).

\bgroup
\def\arraystretch{1.2}
\begin{table*}
\centering
\caption{Source properties, including multi-wavelengh flux densities in the literature, redshift information, as well as derived 1.4 GHz luminosities, SFR, and $q$ values. The SFR and $q^{850{\rm \mu m}}_{1.4 {\rm GHz}}$ value for J123555.14 derived from both the total and AGN-subtracted fractional $L_{{\rm 1.4~GHz}}$, excluding the contribution from the compact AGN core measured by VLBI, are reported in this table.}\label{t1}
\begin{tabular}{lcccccccc}
\hline
\hline
Source Name & $S_{{\rm 1.4 GHz}}$ & $S_{850 {\rm \mu m}}$ & $S_{350 {\rm \mu m}}$ & $z$ & $L_{{\rm 1.4~GHz}}$ & SFR & $q^{850{\rm \mu m}}_{1.4 {\rm GHz}}$ & $q^{350{\rm \mu m}}_{1.4 {\rm GHz}}$ \\
 & [$\mu$Jy] & [mJy] & [mJy] & & [10$^{25}$ W~Hz$^{-1}$] & [10$^4$ \(M_\odot\) $yr^{-1}$] & & \\
\hline
J123555.14 & 212.0 $\pm$ 13.7 & 5.4 $\pm$ 1.9 & 23.1 $\pm$ 0.6 & 1.875 & 0.55 $\pm$ 0.04 & 0.77 $\pm$ 0.06 & 1.41 $\pm$ 0.16 & 2.04 $\pm$ 0.03 \\
J123555.14$^*$ & $-$ & $-$ & $-$ & $-$ & 0.028 $\pm$ 0.002 & 0.039 $\pm$ 0.003 & 2.42 $\pm$ 0.15 & $-$ \\
J123600.10 & 262.0 $\pm$ 17.1 & 6.9 $\pm$ 2.0 & 57.2 $\pm$ 0.6 & 2.710 & 1.66 $\pm$ 0.11 & 2.32 $\pm$ 0.15 & 1.42 $\pm$ 0.13 & 2.34 $\pm$ 0.03 \\
J131225.73 & 752.5 $\pm$ 4.2  & 4.1 $\pm$ 1.3 & < 14.7 & 1.554 & 1.23 $\pm$ 0.01 & 1.72 $\pm$ 0.01 & 0.74 $\pm$ 0.14 & < 1.29 \\
J163650.43 & 221.0 $\pm$ 16.0 & 8.2 $\pm$ 1.7 & 45.9 $\pm$ 2.9 & 2.378 & 1.02 $\pm$ 0.07 & 1.42 $\pm$ 0.10 & 1.57 $\pm$ 0.10 & 2.32 $\pm$ 0.04\\
\hline
\multicolumn{9}{l}{\footnotesize{$^{*}$AGN-subtracted components.}}\\
\end{tabular}
\end{table*}
\egroup

In this paper, we present 1.6 GHz VLBI observations of four SMGs at $z$ $\sim$ 2 using the European VLBI Network (EVN). For a subset of sources we also use e-MERLIN observations from the e-MERlin Galaxy Evolution (eMERGE) Great Observatories Origins Deep-North (GOODS-N) survey (\citealt{muxlow2005}; 2020 in prep.). The sources are SMG~J123555.14+620901.7, SMG~J123600.10+620253.5, SMG~J131225.73+423941.4, and SMG~J163650.43+405734.5. These are located in the HDF-N, SSA-13, and ELAIS-N2 fields and were selected from \citet[][hereafter \citetalias{chapman2005}]{chapman2005}. The four sources that were chosen have $z$ $\sim$ 2, which is the mean redshift of the \citetalias{chapman2005} sample, an epoch that may represent the peak of quasar activity \citep{wolf2003, shankar2009}. On the basis of the results of \citet{chi2013} who detected 12 radio sources brighter than 150 $\mu$Jy in the HDF-N, the targets were also chosen to have a total VLA 1.4 GHz flux density $>$ 200 $\mu$Jy, which is much higher than the average value of $\sim$ 110 $\mu$Jy of the whole sample. The VLA 1.4 GHz luminosities of the four sources are between 10 to 30 times more luminous than the local ultra-luminous starburst Arp 220.

This paper is organised as follows.\ In Section \ref{2} we describe the VLBI observations and the data reduction methodology; in Section \ref{3} we present and discuss our results and the derived source properties compared with other data at different frequencies; and finally, we note the main conclusions of the paper in Section \ref{4}. For this paper, we assume a flat $\Lambda$CDM universe with $H_0$ = 67.8 $\pm$ 0.9 km s$^{-1}$ Mpc, $\Omega_m$ = 0.308 $\pm$ 0.012 and $\Omega_{\Lambda}$ = 0.692 $\pm$ 0.012 \citep{PC2016}.\\

\section{Observations and data analysis} \label{2}

\subsection{EVN observations} \label{2.1}
The source properties, including multi-wavelength flux densities in the literature, redshift information, as well as derived 1.4 GHz luminosities and $q$ values (see Section \ref{3} for details) are listed in Table \ref{t1}. The flux densities at 1.4 GHz (VLA) and at 850 $\mu$m, as well as the redshift information are taken from \citet{chapman2005}. The 350 $\mu$m fluxes are from the \textit{Herschel} point-source catalogue\tablefootnote{http://archives.esac.esa.int/hsa/whsa/} except for the upper limit of J131225.73, which was obtained by \cite{dowell2003} with second-generation Submillimetre High Angular Resolution Camera (SHARC-2) observations. The VLA 1.4 GHz luminosities were derived from their observed 1.4 GHz fluxes without applying any $k$-correction. The SFR, assuming that the radio emission arises purely from star-formation processes, was calculated following \citet{kennicutt1998}:

\begin{equation}
    {\rm SFR} (M_\odot~yr^{-1}) = 1.4 \times 10^{-28}~L_v({\rm ergs~s^{-1}~ Hz^{-1}}),
\end{equation}

where $L_v$ represents the luminosity at a certain frequency $v$; we used the VLA 1.4 GHz radio luminosities to derive the corresponding SFRs. For the VLBI detected source, J123555.14, the SFR values in Table \ref{t1} were derived with and without the contribution from the compact VLBI-detected AGN component. No $k$-correction was applied to the far-IR-radio correlation parameter $q^{850 {\rm \mu m}}_{1.4 {\rm GHz}}$ and $q^{350 {\rm \mu m}}_{1.4 {\rm GHz}}$ values reported here (see Section \ref{3.2} for details).

Our EVN survey was split into two 12-hour observing sessions with project codes EC029A and EC029B, respectively (PI S. Chi). EC029A was observed on 6 November 2009 targeting J131225.73 (SSA-13) and J163650.43 (ELAIS-N2), and EC029B was observed on 7 November 2009 targeting J123555.14 and J123600.10 (both are located in GOODS-N). The eight telescopes listed in Table \ref{t2} were used in both sessions, including the 100 m Effelsberg telescope, the 76 m Lovell telescope, and the 25 m Urumqi telescope in China. The array provided an angular resolution of $\sim$ 10 mas and a 1-$\sigma$ sensitivity of $\sim$ 16 $\mu$Jy beam$^{-1}$.

\bgroup
\def\arraystretch{1.2}
\begin{table}
\centering
\caption{Telescopes used in the observations (ordered alphabetically).\label{t2}}
\begin{tabular}{ccc}
\hline
\hline
Name & Location & Diameter (m)\\
\hline
Effelsberg & Germany & 100\\
Lovell & UK & 76\\
Medicina & Italy & 32\\
Noto & Italy & 32\\
Onsala & Sweden &25\\
Torun & Poland& 32\\ 
Urumqi & China & 25\\
Westerbork & The Netherlands & 25\\
\hline
\end{tabular}
\end{table}
\egroup

\bgroup
\def\arraystretch{1.2}
\begin{table*}[ht!]
\centering
\caption{Information about the sources and calibrators. The coordinates of the undetected sources, J131225.73 and J163650.43, and the calibrators are taken from the references indicated below. For the other sources, the coordinates were measured with AIPS task {\tt\string JMFIT}.\label{t3}}
\begin{tabular}{cccccc}
\hline
\hline
Project & Source Field & Source Name & RA (J2000) & Dec (J2000) & Role\\
\hline
 & SSA-13 & J131225.73 & 13:12:25.734  & +42:39:41.47$^{a}$ & Target\\
 & & J1317+4115 & 13:17:39.1938 & +41:15:45.618$^{b}$ & Phase calibrator for J131225.73\\
EC029A & ELAIS-N2 & J163650.43 & 16:36:50.435 & +40:57:34.46$^{c}$  & Target\\  
 & & J1640+3946 & 16:40:29.6328 & +39:46:46.028$^{b}$ & Phase calibrator for J163650.43\\
  & & 3C345 & 16:42:58.810 & +39:48:36.99$^{b}$ & Fringe finder\\
 \hline
 & GOODS-N & J123555.14 & 12:35:55.1263 & +62:09:01.739 & Target\\
EC029B & GOODS-N & J123600.10 & 12:36:00.0743 & +62:02:53.670  & Target\\
 & & J1241+6020 & 12:41:29.5906 & +60:20:41.322$^{b}$ & Primary calibrator\\
 & & J1234+619 & 12:34:11.7413 & +61:58:32.480 & Secondary calibrator\\
\hline
\multicolumn{6}{l}{\footnotesize{$^{a}$\citet{fomalont2006}}}\\
\multicolumn{6}{l}{\footnotesize{$^{b}$VLBA calibrator catalogue (http://www.vlba.nrao.edu/astro/calib/)}}\\
\multicolumn{6}{l}{\footnotesize{$^{c}$\citet{ivison2002}}}\\
\end{tabular}
\end{table*}
\egroup

The sources were observed in the standard phase referencing mode at 1.6 GHz ($\lambda$ $\approx$ 18cm). For EC029A, we used the bright quasar 3C345 as a 'fringe finder'. Two well-established VLBI calibrators, J1317+4115 ($\sim$ 200 mJy) and J1640+3946 ($\sim$ 890 mJy), lying 1.7 and 1.4 degrees away from the target sources were used as phase calibrators. Each target, along with its phase calibrator, was observed with a cycle time of 10 minutes (8 minutes on the target and 2 minutes on the phase calibrator). The fringe finder was observed for 4 minutes in the middle of the session. For EC029B, we had an $\sim$ 190~mJy primary phase calibrator J1241+6020, which is located $\sim$ 1.5$\degr$ from the centre of the HDF-N, and an $\sim$ 17~mJy secondary phase calibrator J1234+619 lying $\sim$ 25$\arcmin$ away from the HDF-N centre. In this case, the observations cycled between the primary calibrator, the secondary calibrator, J123600.10 (the target), the secondary calibrator, and J123555.14 (the target). Details about the calibrators are listed in Table \ref{t3}. The secondary calibrator of EC029B, J1234+619, was found to be displaced with respect to the correlated position by $\sim$ 20~mas. The improved position measured by these observations via the AIPS task {\tt\string JMFIT} is given in Table \ref{t3}.

Our observations were recorded at 1024 Mbits/sec (Nyquist sampled with two-bit encoding, dual-polarisation, 8 $\times$ 16 MHz IF channels) for a total observing time of $\sim$ 24 hours. Two-second integrations and 16 spectral channels per 16 MHz baseband were adopted for the correlation parameters. Our data were correlated at the Joint Institute for VLBI ERIC \footnote{www.jive.eu/} (JIVE).\\

\subsection{Data analysis} \label{2.2}

The observed data were analysed using the Astronomical Image Processing System (AIPS)\footnote{www.aips.nrao.edu/}. All of the sources and the calibrators were calibrated with the following strategy:
An initial amplitude calibration that was derived from the system temperature and the gain curves of the telescopes (available from the JIVE archive as a calibration table generated by the pipeline processes) was applied first. Thereafter, bad data with abnormally high or corrupted amplitude or phase information were removed using the AIPS tasks {\tt\string SPFLG} and {\tt\string CLIP}. The dispersive delays were then corrected for using an ionospheric map that was implemented within {\tt\string VLBATECR}. The instrumental delays (fixed delay offsets between the IF channels were calibrated by running {\tt\string FRING} on data from a single scan on a strong source (3C345 and J1241+6020 for EC029A and EC029B, respectively). We performed fringe-fitting on the fringe finder and primary phase calibrators to calibrate the phase and phase-rates by also using {\tt\string FRING}. Finally, we performed a bandpass calibration using {\tt\string BPASS} by again employing 3C345 and J1241+6020 for EC029A and EC029B, respectively.

We employed different strategies for EC029A and EC029B for self-calibration. For EC029A, we used {\tt\string SPLIT} to separate the phase-calibrators (J1317+4115 and J1640+3946) and generated the best possible self-calibrated maps of these sources. The clean-components of these maps were then used to update the source model used by {\tt\string FRING}, and the phase and phase-rates were re-determined. After applying the new corrections, we used {\tt\string SPLIT} again to separate the calibrators, then successive loops of {\tt\string IMAGR} and {\tt\string CALIB} resulted in the final maps for these sources. The amplitude and phase corrections derived from {\tt\string CALIB} were then applied to the target sources using {\tt\string CLCAL}. The target data were then separated using {\tt\string SPLIT} and dirty images generated with {\tt\string IMAGR}. 

For EC029B, a similar approach was implemented with corrections from the primary calibrator, which were  determined by {\tt\string FRING} and later refined by {\tt\string CALIB}, including amplitude corrections. These were applied to the secondary calibrator and the targets. The initial images made of the secondary calibrator show it to be shifted about 20 mas from the phase centre (correlated position) of the map. This position was known to be an error but it remained uncorrected at the time the data were correlated. {\tt\string FRING} was re-run on the secondary calibrator using the new map with the correct position, and the phase and phase-rate corrections were also applied to the targets.

Of the four targeted sources, only one was unambiguously detected by VLBI, J123555.14, which is located in the HDF-N. The other three sources were not detected. We generated maps using highly tapered uv-data (excluding the Urumqi telescope), but no further detections were found. We used the Common Astronomy Software Applications (\textsc{casa}; \citealt{mcmullin2007}) task {\tt\string viewer} to generate the radio contour maps of the sources.

\bgroup
\def\arraystretch{1.2}
\begin{table*}[ht!]
\centering
\caption{Derived source properties including the observed EVN peak flux densities, the integrated flux densities, the deconvolved beam sizes, r.m.s. noise levels, the calculated brightness temperatures corrected for redshift by a factor of (1 + $z$), and the recovered VLA flux fractions. For the undetected sources, we derived the 3-$\sigma$ upper limit on the $S_{{\rm VLBI}}/S_{{\rm VLA}}$ ratio. The de-convolved angular sizes of the sources measured by a Gaussian fitting using {\tt\string JMFIT} and the spatial sizes at the distances of the sources calculated using the redshift information \citep{wright2006}. The VLBI detected source J123555.14 has two components, as the southern component was only detected at a 4-$\sigma$ level providing weak reliability, we only present its brighter northern component here. For the three undetected sources, we derived their upper limits with a 3$\sigma$ threshold.\label{t4}}
\begin{tabular}{cccccccc}
\hline
\hline
Source Name & EVN $S_{p}$ & EVN $S_{i}$ & $S_{{\rm VLBI}}/S_{{\rm VLA}}$ & Beam & $T_b$ & Angular Size &
Linear Size\\
 & [$\mu$Jy beam$^{-1}$] & [$\mu$Jy] & & [mas $\times$ mas ($\degr$)] & [10$^5$ K] & [mas $\times$ mas ($\degr$)] & [parsec$^{2}$]\\
\hline
J123555.14 & 110.2 $\pm$ 15.2 & 201.1 $\pm$ 40.2 & 0.95 & 12.8 $\times$ 10.4 (14.5) & 5.2 $\pm$ 0.7 & 23.9 $\times$ 22.3 (131.4) & 116.7 $\times$ 186.9\\
J123600.10 & $\textless$ 42.6 & -- & < 0.16 & 12.6 $\times$ 10.5 (11.0) & $\textless$ 5.6 & -- & -- \\
J131225.73 & $\textless$ 41.0 & -- & < 0.05 & 17.8 $\times$ 11.7 ($-$1.0) & $\textless$ 2.4 & -- & -- \\
J163650.43 & $\textless$ 47.2 & -- & < 0.22 & 18.5 $\times$ 11.6 (1.4) & $\textless$ 3.5 & -- & -- \\
\hline
\end{tabular}
\end{table*}
\egroup

Table \ref{t4} summarises the VLBI results. In this table, the angular size of J123555.14 was measured by fitting a single 2D Gaussian to the source using the AIPS task, {\tt\string JMFIT}. The linear size at the distance of this source was calculated following \citet{wright2006}\footnote{http://www.astro.ucla.edu/~wright/CosmoCalc.html}.\\

\subsection{eMERGE DR1 data} 

Two sources in our sample (J123555.14 and J123600.10) are also part of the eMERGE Data Release 1 (eMERGE DR1). The eMERGE DR-1 dataset provides a very sensitive image of the central $\sim$ 15$\arcmin$ of the GOODS-N field at 1.5 GHz, as observed by e-MERLIN and the Jansky Very Large Array (JVLA).

The source detected by VLBI, J123555.14, lies within the FoV of eMERGE. The VLBI and eMERGE images are presented in Figure~\ref{f2}. The resolution of the eMERGE DR1 maximum sensitivity image is 890 $\times$ 780 milliarcsec$^2$ and the 1-$\sigma$ root mean square (r.m.s.) noise level reaches $\sim$ 1.71 $\mu$Jy beam$^{-1}$ in the central area and $\sim$ 2.37 $\mu$Jy beam$^{-1}$ near the J123555.14 source position. We also created an additional re-weighted eMERGE 1.5 GHz image of J123555.14 with a resolution that matched the published JVLA 5.5 GHz map \citep{guidetti2017}. The primary beam was corrected with a beam size of 560 $\times$ 470 milliarcsec$^2$. It reaches an r.m.s noise level of $\sim$ 1.94 $\mu$Jy beam$^{-1}$ in the central area and $\sim$ 2.54 $\mu$Jy beam$^{-1}$ near the position of J123555.14. The JVLA 5.5 GHz image, which was used along with the re-weighted 1.5 GHz map to derive the spectral index (see Section \ref{3.3}), has an r.m.s noise level of $\sim$ 14.0 $\mu$Jy beam$^{-1}$ near the source position.

J123600.10 lies slightly outside the field of the eMERGE maximum sensitivity map as the FoV is restricted due to the limited extent of the Lovell Telescope primary beam response. We reprocessed the eMERGE data with the Lovell Telescope flagged to make a clear detection (contours shown in Figure \ref{f2}b). The reprocessed image has less sensitivity than the central DR1 image, but it  still reaches an r.m.s. noise level of $\sim$ 1.77 $\mu$Jy beam$^{-1}$ near the source position with a resolution of 770 $\times$ 750 milliarcsec$^2$ and goes significantly deeper than our VLBI images. As J123600.10 lies near the edge of this image, there is some bandwidth smearing of the source at the level of $\sim$ 10\%. \\


\section{Results and discussion} \label{3}


\subsection{Brightness temperature\label{3.1}}

The brightness temperature ($T_b$) of a source with a redshift $z$ is given by:

\begin{equation}
T_b  =  1.22 \times 10^{12}(1 + z)\left(\frac{S_{\nu}}{1 {\rm Jy}}\right)\left(\frac{\nu}{1 {\rm GHz}}\right)^{-2}\left(\frac{\theta_{maj}\theta_{min}}{1 {\rm mas^2}}\right)^{-1} {\rm K},
\end{equation}
where $S_{\nu}$ is the peak brightness and $\nu$ is the observing frequency; $\theta_{maj}$ and $\theta_{min}$ denote the deconvolved major and minor axes of the source \citep{condon1982, ulvestad2005}. For the three undetected sources, we derived $5\sigma$ upper limits for their brightness temperatures (see Table \ref{t4}).

A well-established upper limit to the brightness temperature of 10$^5$ K exists in star-forming systems \citep{condon1992,lonsdale1993}. In the local Universe, such a brightness temperature can be achieved by either AGN or supernovae activity \citep{kewley2000}. However, at a more distant universe ($z > 0.1$), where the VLBI detection threshold exceeds what stellar non-thermal sources can reach, this value can typically only be attained by AGN \citep{middelberg2010,middelberg2013}. Notably, while one can be confident of a compact, approximate milli-arcsecond source with $T_b$ $>$ 10$^5$ K being an AGN, it is not certain that AGN in sources below this criteria can be ruled out. Those sources may contain extended structures, such as jets and lobes, and only a smaller fraction of their flux is from the compact central core \citep{jarvis2019, muxlow2005}.

The detected source J123555.14 has a brightness temperature of 5.2 $\pm$ 0.7 $\times$ 10$^5$ K, which is approximately five times higher than the star-forming envelope. This supports the interpretation that J123555.14 contains an AGN core. The undetected sources, however, have upper limits on their brightness temperatures exceeding $10^5$ K, which cannot completely rule out AGN activity in these objects.\\

\subsection{Infrared-radio correlation\label{3.2}}

\begin{figure*}[!htb]
\centering
\includegraphics[width=12cm]{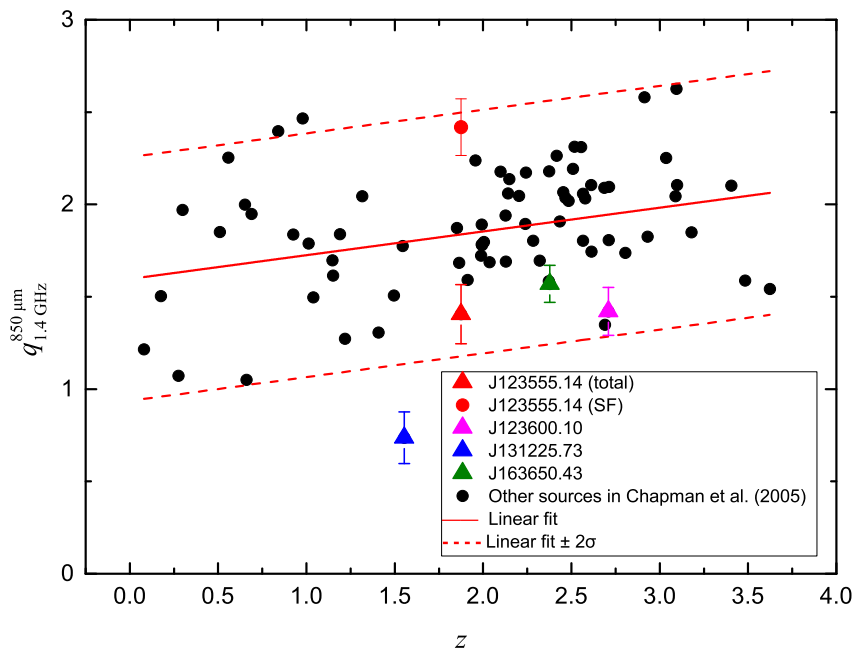}
\caption{The distribution of non-k-corrected $q^{850 {\rm \mu m}}_{1.4 {\rm GHz}}$ versus redshift of 76 sources in the \citetalias{chapman2005} sample, including the four VLBI observed sources presented in this paper. The solid red line is a linear fit on the whole sample, and the dashed lines constrain the region within a 2-$\sigma$ dispersion to the fit ($\pm$ 0.66). The VLBI observed sources are labelled individually with errors. The red triangle represents the $q^{850 {\rm \mu m}}_{1.4 {\rm GHz}}$ value of J123555.14, which was calculated with its full 1.4 GHz flux density, while the red dot represents its $q^{850 {\rm \mu m}}_{1.4 {\rm GHz}}$ value associated with purely star-formation processes, which is higher than the majority of the sample. J131225.73 (blue triangle) is clearly an outlier in the plot. \label{f1}\\}
\end{figure*}

Our four sources are part of a larger sample of SMGs from \cite{chapman2005} that have good positions and measured redshifts. Table \ref{t1} presents the source flux densities at 1.4 GHz and 850 $\mu$m provided in \citet{chapman2005} and the derived $q^{850 {\rm \mu m}}_{1.4 {\rm GHz}}$ values. We define the latter as:

\begin{equation}
    q^{850{\rm \mu m}}_{1.4{\rm GHz}} = \log_{10}(\frac{L_{850{\rm \mu m}}}{L_{{\rm 1.4GHz}}})
,\end{equation}

where $L_{850{\rm \mu m}}$ and $L_{1.4~{\rm GHz}}$ are the luminosities at $850~\mu$m and 1.4 GHz. We also calculated the $q^{350 {\rm \mu m}}_{1.4 {\rm GHz}}$ values for the sources using the same strategy (Table \ref{1}).

A fairly tight far-infrared-radio correlation applies to a wide-range of galaxy types in the local Universe (see \citealt{solarz2019} and references therein). The relation also appears to hold at cosmological distances (e.g. \citet{garrett2002} and \cite{elbaz2002}). Radio-loud AGNs are observed to have much lower values of $q$ on average, compared to radio quiet or star-forming systems. The value of $q$ can therefore help to distinguish between AGN activity and star-forming processes in extragalactic systems (e.g. \citet{condon2002, sargent2012, delhaize2017}).

Figure \ref{f1} presents a plot of $q^{850 {\rm \mu m}}_{1.4 {\rm GHz}}$ versus the redshift for the 76 SMG sources in \citet{chapman2005}, including the four sources observed by VLBI. Since we were mostly interested in seeing whether our sources deviated from the rest of the sample, a simple linear fit was applied to the observed band flux ratios without the application of any $k$-correction. The \citetalias{chapman2005} sample has a mean $q^{850 {\rm \mu m}}_{1.4 {\rm GHz}}$ of $\sim$ 1.85 with a standard deviation from the linear fit of 0.35. Three of our four sources appear to follow the FIR-radio correlation, which is represented by $q^{850 {\rm \mu m}}_{1.4 {\rm GHz}}$ with offsets that are smaller than 2$\sigma$ from the linear fit. 

It is interesting, however, that all three objects are below the correlation, thus, on average, they have larger radio luminosities than expected. However, for the detected source J123555.14, the $q^{850 {\rm \mu m}}_{1.4 {\rm GHz}}$ value increases to 2.42 $\pm$ 0.15 when the contribution of the AGN core is subtracted from the 1.4 GHz radio luminosity. This value is actually higher than most of the sources in Figure \ref{f1}. Considering that the linear fit was performed on a sample of SMG, which is possibly contaminated by AGN, the correlation shown in Figure \ref{f1} might be displaced towards lower q values with respect to that of a sample consisting purely of star-forming galaxies. 

One of the sources, J131225.73, has a value of $q^{850 {\rm \mu m}}_{1.4 {\rm GHz}} \sim$ 0.74, which departs from the mean linear fit by $\sim$ 3$\sigma$. A further discussion on this point is given in Section~\ref{3.3}.

While $q^{850 {\rm \mu m}}_{1.4 {\rm GHz}}$ and $q^{350 {\rm \mu m}}_{1.4 {\rm GHz}}$ are more sensitive to the dust emission measurement, other $q$-values with wider IR waveband coverage, such as $q_{{\rm IR}}$ and $q_L$, are more sensitive to the actual IR emission from the source. We note that $q_{IR}$ is defined as the logarithmic ratio of the rest-frame 8-1000 $\mu$m flux and 1.4 GHz radio flux. \cite{bell2003} measured a median $q_{IR}$ value of 2.64 $\pm$ 0.02 over 164 SMGs, showing no signs of AGN. This value is similar to the medium $q_{{\rm IR}}$ value of 2.59 $\pm$ 0.05 in \cite{thomson2014} involving 76 SMGs. \cite{DM2013} classified sources with $q_{{\rm IR}}$ $<$ 1.68 as radio excess AGNs.\ Additionally, $q_L$ is the logarithmic ratio of the far-IR and 1.4 GHz radio luminosity, where the far-IR luminosity is estimated by the flux in a wide band centred at 80 $\mu$m. \citet{kovacs2006} derived an average $q_L$ value of 2.14 with an intrinsic spread of 0.12. Sources with a $q_L$ value that is more than 2$\sigma$ lower than the mean value are likely to be hosts of a radio-loud AGN. In Section \ref{3.3}, we use the $q$ values of the sources from the literature to support our arguments. Both the $q_{{\rm IR}}$ value of J123555.14 and the $q_L$ values of J131225.73 and J163650.43 are below the mean values mentioned above, which is in agreement with the $q^{850 {\rm \mu m}}_{1.4 {\rm GHz}}$ distribution characteristics of Figure \ref{f1}.\\


\subsection{Source description\label{3.3}}

\begin{figure*}[ht!]
  \centering
  \begin{tabular}[c]{cc}
    \begin{subfigure}[b]{0.48\textwidth}
      \includegraphics[width=\textwidth]{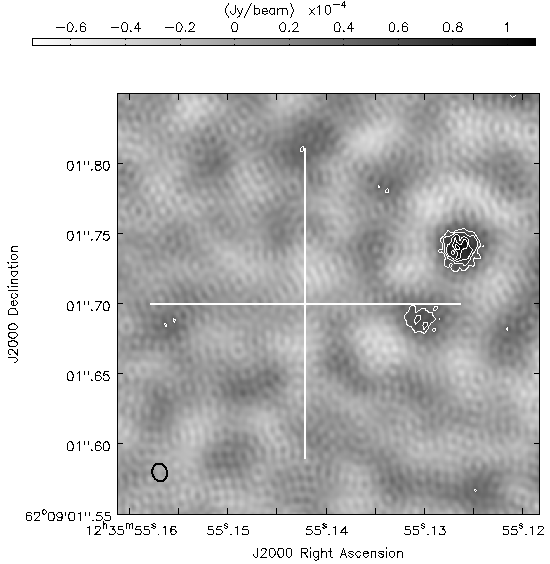}
      \caption{ }
    \end{subfigure}&
    \begin{subfigure}[b]{0.48\textwidth}
      \includegraphics[width=\textwidth]{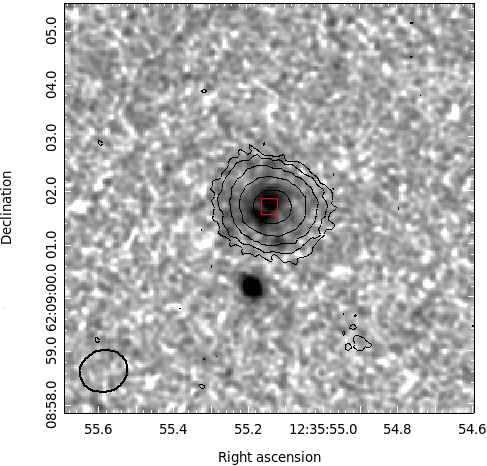}
      \caption{ }
      \label{fig:gull2}
    \end{subfigure}\\
    \begin{subfigure}[b]{0.48\textwidth}
      \includegraphics[width=\textwidth]{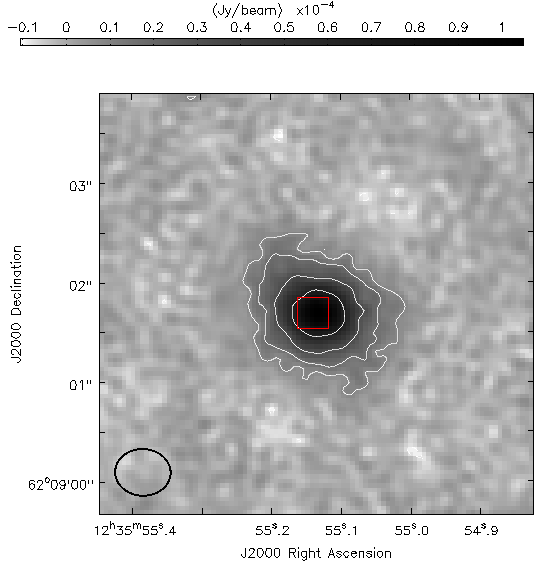}
      \caption{ }
      \label{fig:tiger}
    \end{subfigure}&
    \begin{subfigure}[b]{0.48\textwidth}
      \includegraphics[width=\textwidth]{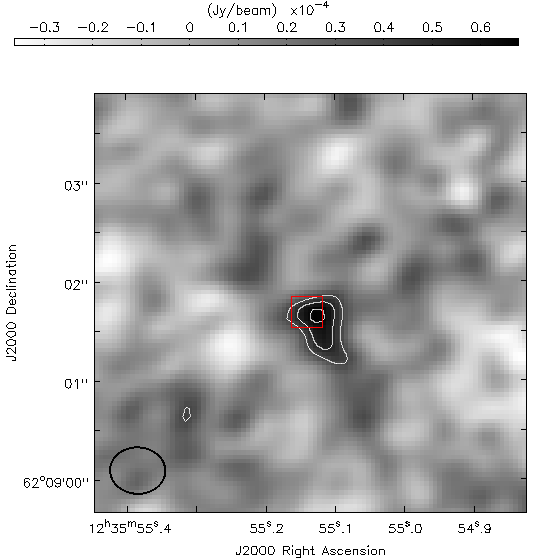}
      \caption{ }
    \end{subfigure}
  \end{tabular}
  \caption{Contoured maps for the VLBI-detected source J123555.14+620901.7. (a) EVN 1.6 GHz image plotted with contour levels (3, 4, 5, 6, and 7) $\times$ the r.m.s noise level; the white cross presents the VLA coordinates measured by \citet{richards2000}. The FoV of this image is shown by a red box in each of the other three sub-figures. (b) The $Hubble$ CANDELS F814W ACS image in which the source shows a face-on disc-like morphology with a close companion for clarity; the image was smoothed with a 2D Gaussian convolution with FWHP of $\sim$ 0.1$\arcsec$. The overlaid contour was plotted at 3, 6, 12, 24, and 48 $\times$ the r.m.s noise of the eMERGE-JVLA maximum sensitivity image at 1.5 GHz. (c) The eMERGE re-weighted 1.5 GHz map presented with contour levels 3, 6, 12, and 24 $\times$ the r.m.s noise level. (d) The eMERGE 5.5 GHz map plotted with contour levels (3, 4, and 5) $\times$ the r.m.s noise level. The detected compact AGN core is clearly shown in the images. In each panel, the beam pattern of the contours is illustrated with a black ellipse.\label{f2}}
\end{figure*}

\begin{figure*}[ht]
  \centering
  \begin{tabular}[c]{cc}
    \begin{subfigure}[b]{0.48\textwidth}
      \includegraphics[width=\textwidth]{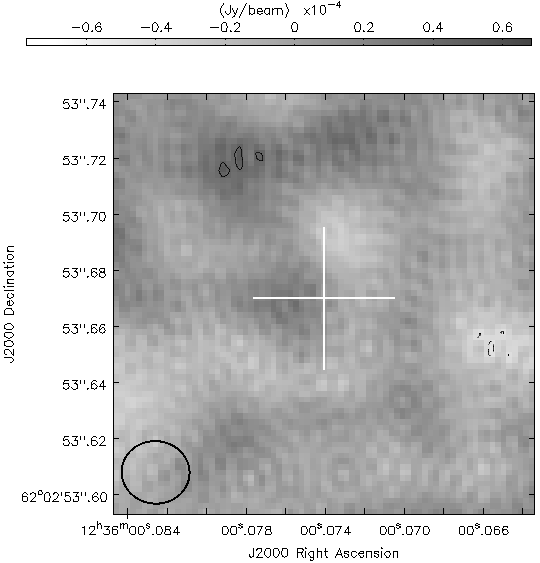}
      \caption{ }
    \end{subfigure}&
    \begin{subfigure}[b]{0.48\textwidth}
      \includegraphics[width=\textwidth]{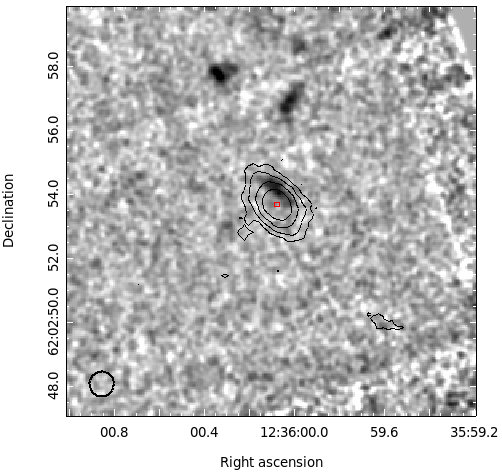}
      \caption{ }
    \end{subfigure}\\
  \end{tabular}
\caption{(a) EVN 1.6 GHz VLBI image of the field associated with J123600.10 (undetected). The white cross on the map presents the position of the source as measured in the re-processed eMERGE image. The map size corresponds to the 2$\sigma$ uncertainties in the VLA position. This image was plotted with contour levels $-$3 and 3 $\times$ the r.m.s noise level. (b) $Hubble$ NICMOS NIC2 F160W image of J123600 in which the source shows a disc-like structure with close projected companions.\ For clarity, the image was smoothed with a 2D Gaussian convolution with a FWHP of $\sim$ 0.15$\arcsec$. The overlaid contour was plotted at 3, 6, 12, and 24 $\times$ the r.m.s noise level of the eMERGE-JVLA moderate resolution image at 1.5 GHz. The red box shows the FoV of (a). As shown in (b), the radio and optical images have similar morphologies with reasonable position offset of $\sim$ 0.7$\arcsec$. In each panel, the \textbf{beam} pattern of the contours is illustrated with a black ellipse. \label{f3}\\}
\end{figure*}

\begin{figure*}[hbt!]
  \centering
  \begin{tabular}[c]{cc}
    \begin{subfigure}[b]{0.48\textwidth}
      \includegraphics[width=\textwidth]{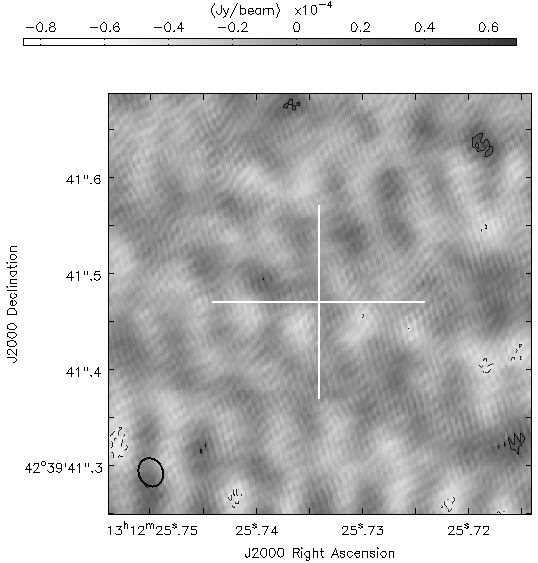}
      \caption{ }
    \end{subfigure}&
    \begin{subfigure}[b]{0.48\textwidth}
      \includegraphics[width=\textwidth]{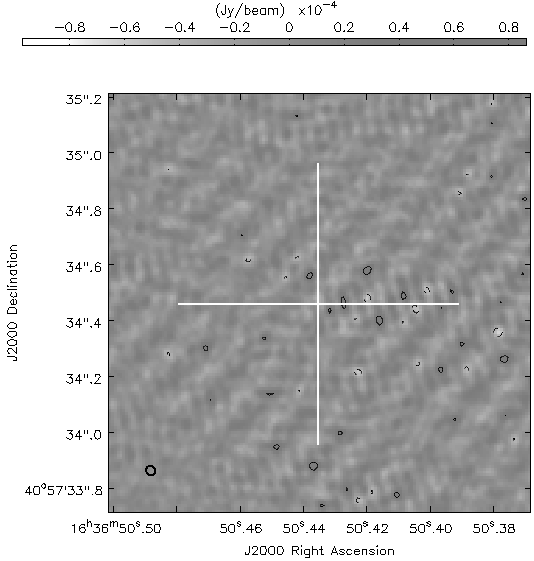}
      \caption{ }
    \end{subfigure}\\
  \end{tabular}
\caption{EVN 1.6 GHz images of (a) J131225.73 (undetected) and (b) J163650.43 (also undetected). The expected positions and errors in the maps represented by the crosses are provided in \citet{fomalont2006} and \citet{ivison2002} for (a) and (b), respectively. The maps are presented with contour levels $-$3 and 3 $\times$ the r.m.s noise. The white crosses on the maps are centred on the expected positions of the sources. For clarity, the map sizes correspond to the 2$\sigma$ and 1.5$\sigma$ uncertainties in the VLA positions for (a) and (b), respectively. In each panel the VLBI beam pattern is illustrated with a black ellipse.\label{f4}\\}
\end{figure*}

The various contour maps of our four targets are shown in Figures \ref{f2}, \ref{f3}, and \ref{f4}. The crosses on the figures are centred on their expected a priori positions and scaled with respect to the associated error. In Figure \ref{f2}a, the white cross presents the VLA-measured coordinates of J123555.14 given in \citet{richards2000}, the red dotted cross shows the measured position in the eMERGE maximum sensitivity image and the blue one represents the position in the eMERGE 5.5 GHz map. The cross in Figure \ref{f3}a represents the expected position of J123600.10 measured in the re-processed eMERGE image. For the other two undetected sources in Figure \ref{f4}, the expected positions and errors are specified in \citet{chapman2005, fomalont2006, ivison2002} and refer to 1.4~GHz VLA observations.\\

\textbf{J123555.14} was detected and resolved in our VLBI observations. This source is located at a spectroscopic redshift of 1.875 \citep{chapman2005}. Figure~\ref{f2} shows the VLBI, the $Hubble$ optical image, and e-MERLIN DR1 maps of the source. The EVN image shows two components separated by 56.5 mas. The brighter of the two components is consistent with the source detected by \citet{radcliffe2018}. They did not detect the fainter southern component, but we note that this is only detectable in our image at the 4-$\sigma$ level, which has a weak reliability. Therefore, we do not discuss this component in this paper. The brighter northern component was detected at the $\sim$ 7-$\sigma$ level with a brightness temperature of $5.2$ $\pm$ 07 $\times$ 10$^{5}$ K. 

The source shows a face-on disc-like morphology with a close companion ($\sim$ 10 $-$ 20 kpc in projection) in the rest-frame near-UV image (Figure \ref{f2}b). This image was taken by the Advanced Camera for Surveys (ACS) of the \textit{Hubble~Space~Telescope (HST)} with the F814W filter in the Cosmic Assembly Near-infrared Deep Extragalactic Legacy Survey (CANDELS; \citealt{candels2011}) and was retrieved from the $Hubble$ Legacy Archive\footnote{https://hla.stsci.edu/}. The \textit{HST} image was processed by SAO Image DS9\footnote{http://ds9.si.edu/site/Home.html}. For clarity, the image was smoothed with a 2D Gaussian convolution with a FWHP of $\sim$ 0.1$\arcsec$. 

\begin{figure*}[!htb]
\centering
\includegraphics[width=14cm]{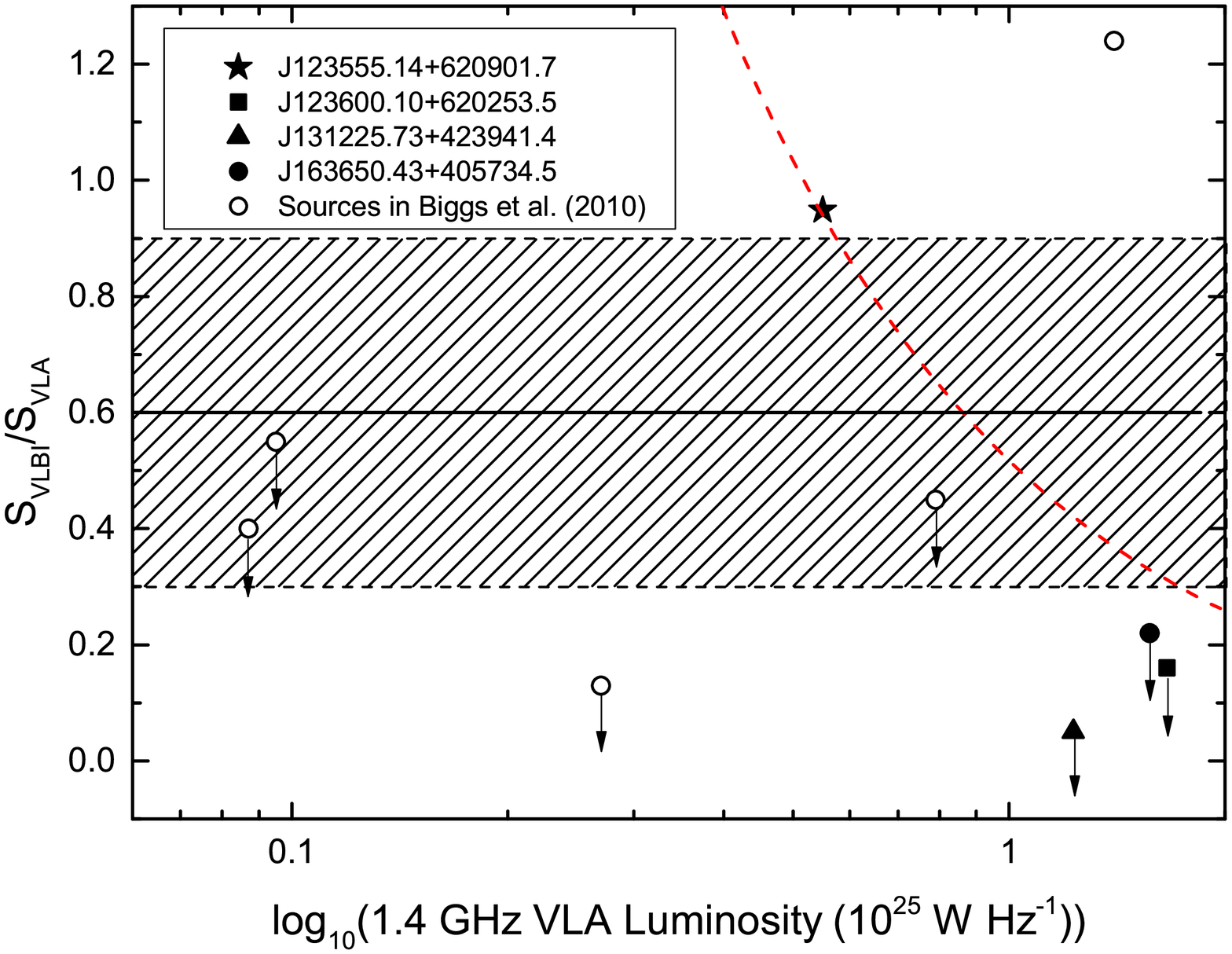}
\caption{VLBI(VLBA)-to-VLA flux ratio ($R$) versus the 1.4 GHz VLA luminosity for the four VLBI observed sources as well as five sources in \citet{biggs2010} with redshift information, of which one in the upper right panel was classified as an AGN. The average $R$ value over $\sim$ 500 VLBA detected sources in \citet{HR2017} of $\sim$ 0.6 is labelled by the horizontal solid line, and the 1-$\sigma$ constraint of $\pm$ 0.3 to the mean value is represented by the shaded area. The VLBI detected source J123555.14 has a high $R$ value of 0.95, showing that a large fraction of its radio emission comes from a compact AGN core. For the VLBI undetected sources, the 3-$\sigma$ upper limits on their $R$ values are presented by downward arrows, which are more than $-1\sigma$ away from the mean value. The red dashed curve indicates where $L_{{\rm VLBI}}$ = 5.5 $\times$ 10$^{24}$ W Hz$^{-1}$ $-$ the 1.6 GHz VLBI luminosity of the detected source J123555.14 $-$ although J123555.14 has only medium VLA luminosity in this figure, its VLBI luminosity is higher than most of the sources. \label{f5}}
\end{figure*}

The \textit{HST} CANDELS image was contoured with the 1.5 GHz eMERGE DR1 maximum sensitivity map, where the source is detected with a peak and integrated flux density of 185.2 $\pm$ 4.6 $\mu$Jy. This value is consistent with the VLBI flux within the errors. Moreover, this source has a total 1.4 GHz VLA flux density of 212.0 $\pm$ 13.7~$\mu$Jy, which corresponds to a radio luminosity of 5.50 $\pm$ 0.4 $\times$ 10$^{24}$ W Hz$^{-1}$. As shown in Figure \ref{f5}, this source has a prominent VLBI-to-VLA flux density ratio of 0.95, which is much higher than the mean value of $\sim$ 0.6 measured by \citet{HR2017} over a larger sample of Very Long Baseline Array (VLBA) detected sources, which were observed with a similar, approximate milli-arcsecond resolution and an $\sim$ 10 $\mu$Jy sensitivity. This suggests that most of the radio emission is associated with a compact milli-arcsecond central core. Another possible explanation for a large value of $S_{{\rm VLBI}}/S_{{\rm VLA}}$ is source variability. In particular, the VLBI and VLA data in this paper were taken at different epochs. Nevertheless, only a few percent of faint radio sources are expected to be variable and this particular source was observed as being variable in the recent study of \citet{radcliffe2019}. The source has a SFR of 0.77 $\pm$ 0.06 $\times$ 10$^{4}$ \(M_\odot\) $yr^{-1}$, which was calculated with the total $L_{1.4~{\rm GHz}}$. In removing the 95\% VLBI recovered luminosity which is considered to be associated with an AGN, a corresponding SFR of 385 $\pm$ 30 \(M_\odot\) $yr^{-1}$ is indicated by the 1.4 GHz radio luminosity contributed by star-forming processes (0.28 $\pm$ 0.02 $\times$ 10$^{24}$ W Hz$^{-1}$). 


The source appears to be slightly resolved in the eMERGE 1.5 and 5.5~GHz images, the latter has a peak brightness of $\sim$ 5$\sigma$. The total flux densities of the eMERGE 1.5 and 5.5~GHz images yield a spectral index of $\sim$ $-$0.49. This relatively flat spectral index is indicative of an AGN component being the dominant source of radio emission. This source was also confirmed to be an AGN in \citet{radcliffe2018}. Its mid-IR spectrum also shows evidence of an AGN in this source \citep{pope2008}. \citet{hainline2011} suggest that an estimated fraction of 0.7 $\sim$ 0.8 of its \textit{Spitzer}-Infrared Array Camera (IRAC) 8 $\mu$m emission was contributed by a power-law component ($f_{\lambda}$ $\sim$ $\lambda^{\alpha}$ where $\alpha$ = 3 was the best fit), which is considered to likely originate from an AGN. By fitting local spectral energy distributions (SEDs) with 24 $\mu$m \textit{Spitzer} photometry, \citet{murphy2009} suggested that more than 50\% of its 8-1000 $\mu$m total IR energy budget could be contributed by AGN activity. This source has a S$_{850{\rm \mu m}}$ and S$_{350{\rm \mu m}}$ of 5.4 $\pm$ 1.9 mJy and 23.1 $\pm$ 0.6 mJy; this yields a $q^{850{\rm \mu m}}_{1.4 {\rm GHz}}$ and $q^{350{\rm \mu m}}_{1.4 {\rm GHz}}$ of 1.41 $\pm$ 0.16 and 2.04 $\pm$ 0.03, respectively (see Table \ref{t1}). Although the $q^{850{\rm \mu m}}_{1.4 {\rm GHz}}$ and $q^{350{\rm \mu m}}_{1.4 {\rm GHz}}$ values of this source seem to follow the FIR-radio correlation, it has a $q_{{\rm IR}}$ of 1.51 \citep{murphy2009}. This value is $\sim$ 4$\sigma$ lower than the median $q_{{\rm IR}}$ value of $\sim$ 2.60 \citep{bell2003, thomson2014}. According to the argument that \citet{DM2013} raise that sources with $q_{{\rm IR}}$ $<$ 1.68 are likely radio excess AGNs, the low $q_{{\rm IR}}$ value of this source is a strong indicator of an AGN. After subtracting the AGN-contributed radio luminosity, the source has a $q^{850{\rm \mu m}}_{1.4 {\rm GHz}}$ and $q^{350{\rm \mu m}}_{1.4 {\rm GHz}}$ values of 2.42 $\pm$ 0.15 and 3.05 $\pm$ 0.02, respectively.\ These values may reflect the $q$ values for purely star-forming systems. As shown in Figure \ref{f1}, the star-formation-associated value of $q^{850{\rm \mu m}}_{1.4 {\rm GHz}}$ is higher than the majority of the \citetalias{chapman2005} SMG sample. The source also shows evidence of an AGN at X-ray wavelengths, as suggested by \citet{alexander2005} and \citet{DM2013}. \\





\textbf{J123600.10} was  not detected in our VLBI observations with a 3-$\sigma$ brightness temperature limit of $< 5.6 \times 10^5$ K and it is located at a redshift of 2.710 \citep{chapman2005}. This source is detectable in the eMERGE re-weighted image with an angular size of $859 \times 253$ mas$^2$, and the measured peak and integrated flux densities are 52.4 $\pm$ 1.7 $\mu$Jy beam$^{-1}$ and 83.5 $\pm$ 4.1 $\mu$Jy, respectively. This yields a brightness temperature of $\sim$ 523 K. Since the peak brightness measured in the eMERGE re-weighted image is only approximately five times the VLBI detecting threshold, it is not surprising that we did not detect this source with VLBI.

The contours of the eMERGE re-weighted image for this source are shown in Figure \ref{f3}b on top of the \textit{HST} image taken with the Near Infrared Camera and Multi-Object Spectrometer (NICMOS) using the NIC2 camera and the F160W filter in \citet{swinbank2010} (project ID: 9506). The source shows a disc-like morphology with close companions. The $Hubble$ NICMOS image was retrieved from the $Hubble$ Legacy Archive and was processed by SAO Image DS9.\ For clarity, the image was smoothed with a 2D Gaussian convolution with a FWHP of $\sim$ 0.15$\arcsec$.
 
This source has a S$_{850{\rm \mu m}}$ and S$_{350{\rm \mu m}}$ of 6.9 $\pm$ 2.0 mJy and 57.2 $\pm$ 0.6 mJy; this yields a $q^{850{\rm \mu m}}_{1.4 {\rm GHz}}$ and $q^{350{\rm \mu m}}_{1.4 {\rm GHz}}$ of 1.42 $\pm$ 0.13 and 2.34 $\pm$ 0.03, respectively (Table \ref{t1}).\ Additionally, both follow the FIR-radio correlation. The source has a low VLBI-to-VLA flux ratio upper limit of 0.16, suggesting that a significant fraction of the radio emission is extended and probably associated with star-formation processes. This is supported by other measurements $-$ \citet{chapman2003} suggest that this galaxy is possibly an edge-on merger from its optical morphology. However, this source has a total 1.4 GHz VLA flux density of 262.0 $\pm$ 17.1~$\mu$Jy, which corresponds to a radio luminosity of 1.66 $\pm$ 0.11 $\times$ 10$^{25}$ W Hz$^{-1}$. This value can only be reached by an AGN or extremely strong star-forming galaxies with SFR of 2.32 $\pm$ 0.15 $\times$ 10$^4$ \(M_\odot\) $yr^{-1}$. The high luminosity indicates that extended jet emission from a jet associated with quasar activity may exist in this source \citep{jarvis2019, muxlow2005}, which was undetected in these VLBI observations. \\


\textbf{J131225.73} was not detected in our VLBI observations with a 3-$\sigma$ brightness temperature limit of $< 2.4~\times$ 10$^{5}$ K and it is located at a redshift of 1.554 \citep{chapman2005}. A point source was detected at 8.4~GHz by the VLA at a resolution of 6$\arcsec$ \citep{fomalont2002} with a total flux density of $200.3 \pm 6.5~\mu $Jy. The implied spectral index of the source is relatively steep: $\alpha \sim$ -0.7. The non-detection of the source on VLBI scales and the very low upper limit of the $S_{{\rm VLBI}}/S_{{\rm VLA}}$ ratio of 0.05 may suggest that a significant fraction of the radio emission is associated with star-formation processes.

However, although this is a steep spectrum radio source, it does have a significant radio-excess with relatively low $q$ values. This source has a $S_{850 {\rm \mu m}}$ of 4.1 $\pm$ 1.3 mJy (see Table \ref{t1}), and it is clearly identified as an outlier in Figure \ref{f1}. Indeed, it has the lowest value of $q^{850 {\rm \mu m}}_{1.4 {\rm GHz}}$ out of all of the sources in the \citetalias{chapman2005} sample. This source was undetected with SHARC-2 \citep{dowell2003}; this yields an upper limit to its $S_{350 {\rm \mu m}}$ of 14.7 mJy \citep{laurent2006, kovacs2006}, which corresponds to a $q^{350 {\rm \mu m}}_{1.4 {\rm GHz}}$ of $\lesssim$ 1.29. Moreover, \citet{kovacs2006} derived a $q_L$ value of 0.79 $\pm$ 0.35 for this source; this value is exceptionally lower than their average value of 2.14, thus they suggested that this source likely hosts a radio-loud AGN. In addition, the source is unresolved by the VLA at 1.4 and 8.4~GHz. It has a flux density in excess of 750 $\mu$Jy at 1.4 GHz and a corresponding radio luminosity of 1.23 $\pm$ 0.01 $\times$ 10$^{25}$ W Hz$^{-1}$. Assuming all the radio emission is associated with star-formation processes, an extremely high SFR of 1.72 $\pm$ 0.01 $\times$ 10$^4$ \(M_\odot\) $yr^{-1}$ is implied. 

Given the main observational properties of J131225.73, it is rather surprising that this source goes undetected by the EVN at 1.6~GHz. One possible explanation is that while the radio emission is indeed associated with an AGN, it is extended in nature, which is possibly due to the presence of extended jet features that dominate the total flux density of the source and extend spatially over kiloparsec scales. Since the morphology of the faint radio source population is still largely unknown on these scales, it is possible that VLBI misses (or resolves) many radio AGN that are dominated by extended jet components. \\

\textbf{J163650.43} was not detected in our VLBI observations with an implied 3-$\sigma$ brightness temperature limit of $< 3.5$ $\times$ 10$^{5}$ K and it is located at a redshift of 2.378 \citep{chapman2005}. This source has a S$_{850{\rm \mu m}}$ and S$_{350{\rm \mu m}}$ of 8.2 $\pm$ 1.7 mJy and 45.9 $\pm$ 2.9 mJy; this yields a $q^{850{\rm \mu m}}_{1.4 {\rm GHz}}$ and $q^{350{\rm \mu m}}_{1.4 {\rm GHz}}$ of 1.57 $\pm$ 0.10 and 2.32 $\pm$ 0.04, respectively (see Table \ref{t1}).\ Additionally, both seem to follow the FIR-radio correlation. The source recovers $<$ 22\% of its 1.4 GHz flux; the non-detection on VLBI scales may suggest that a significant fraction of the radio emission is associated with star-formation processes. This is supported by other measurements $-$ \citet{engel2010} classify this source as a close binary galaxy merger because the CO(3-2) and CO(7-6) data show two peaks separated by $\sim$ 3 kpc. A $q_L$ value of 1.75 $\pm$ 0.17 for this sources was measured by \citet{kovacs2006}, this value is slightly lower than the 2-$\sigma$ constraint from the mean value of 2.14, and it was considered to be consistent with the far-infrared to radio correlation. 

However, J163650.43 has a total 1.4 GHz VLA flux density of 221.0 $\pm$ 16.0~$\mu$Jy corresponding to a luminosity of 1.02 $\pm$ 0.07 $\times$ 10$^{25}$ W Hz$^{-1}$. This value can only be reached by an AGN or extremely powerful star-forming galaxies with a SFR of 1.42 $\pm$ 0.10 $\times$ 10$^4$ \(M_\odot\) $yr^{-1}$. The high luminosity suggests that similar to 131225.73, this source may host extended radio jet emission associated with an AGN \citep{jarvis2019, muxlow2005}. Moreover, this source was classified as an AGN-hosting galaxy in \citet{swinbank2004} because of its H$\alpha$ emission with a large line width of 1753 km~s$^{-1}$ in the near-infrared. \citet{hainline2011} suggest that an estimated fraction of 0.6 $\sim$ 0.8 of its \textit{Spitzer}-Infrared Array Camera (IRAC) 8 $\mu$m emission was contributed by a power-law component ($f_{\lambda}$ $\sim$ $\lambda^{\alpha}$ where $\alpha$ = 3 was the best fit), which is considered to likely originate from an AGN.\\



\section{Summary} \label{4}
We have conducted EVN 1.6~GHz observations of four SMGs (J123555.14, J123600.10, J131225.73, and J163650.43,). Out of the four targets, we detected J123555.14 once, which is a source located in the GOODS-N field with a brightness temperature of 5.2 $\pm$ 0.7 $\times$ 10$^{5}$ K. This value exceeds the maximum brightness temperature of $\sim$ 10$^{5}$ K that a star-forming galaxy is expected to reach. We therefore suggest that the radio emission associated with J123555.14 largely arises from AGN processes. We also derived brightness temperature upper limits of the three undetected sources. The non-detections may suggest that most of their radio emission is powered by star-forming processes or extended radio jets. Notably, the three undetected sources all have relatively high 1.4 GHz radio luminosities and/or show evidence of an AGN in other wavebands. This fact may suggest that while SGMs are potential hosts of AGN, star-forming processes and AGN activity probably exist in such systems concordantly. 

In particular, at least one of the sources, J131225.73, shows multi-wavelength properties that would lead us to expect it to be detected by VLBI. It is highly luminous in the 752.5 $\pm$ 4.2 $\mu$Jy or 1.23 $\pm$ 0.01 $\times$ 10$^{25}$ W Hz$^{-1}$ at 1.4 GHz, has an exceptionally low value of $q$, and is unresolved in both the 1.4 and 8.4~GHz VLA observations. We suggest that this source, might be associated with an AGN that is dominated by extended jet emission. It would be interesting to observe the structure of this radio source with intermediate resolution, such as e-MERLIN at 5~GHz, in order to better understand the nature of the source. Since the morphology of the faint radio source population is still largely unknown on these scales, it is possible that VLBI misses (or resolves) many extended radio AGN of this type. As illustrated by the upper limits of the VLBI-to-VLA flux ratio of the un-detected sources, a bright radio source ($\sim$ 200$\mu$Jy at 1.4 GHz) with less than $\sim$ 20\% of it radio flux contributed by an AGN would probably be missed by VLBI observations at a 1-$\sigma$ sensitivity of $\sim$ 10 $\mu$Jy. Surveys with multiple resolution (e.g. VLA, e-MERLIN, and VLBI) are therefore needed to determine what fraction of the extragalactic objects have similar properties and thus would probably be missed by VLBI.\\

\section*{Acknowledgments}
The EVN observations were originally proposed by S. Chi et al. in 2009. Seungyoup Chi passed away in 2011, the final year of his PhD studies. We very much appreciate his contribution to this project, including earlier research on the topic. 
HC is funded by the Science and Technology Facilities Council (STFC) and the China Scholarship Council (CSC) (File No.201704910999), we are thankful for their support.
DMA and IRS acknowledge STFC through grant code ST/P000541/1.
IP acknowledges support from INAF under the PRIN SKA/CTA `FORECaST' project.
JFR is funded by the South Africa Radio Astronomy Observatory and is grateful for their support
This research made use of Astropy, a community-developed core Python
package for Astronomy \citep{astropy2013,astropy2018}.
The European VLBI Network is a joint facility of European, Chinese, South African, and other radio astronomy institutes funded by their national research councils.
e-MERLIN is a National Facility operated by the University of Manchester at Jodrell Bank Observatory on behalf of STFC.
The National Radio Astronomy Observatory is a facility of the National Science Foundation operated under cooperative agreement by Associated Universities, Inc.\\

\bibliographystyle{aa} 
\bibliography{ref}


\end{document}